\documentstyle{article}

\begin{document}

\def\levelset{{\cal X}}
\def\fxx{{\bf x}}
\def\fxy{{\bf y}}
\def\real{I\!\!R}

\begin{center}
{\large {\bf Absoluteness of Velocity Caused by Accelerating Process }}\\
{\large {\bf and Absolute Space-time Theory with Variable Scales }}\\
{\large {\bf  -----The third logically consistent and really rational space-time theory }}\\
\vskip 0.2in
{\large Mei Xiaochun} \\
\vskip 0.2in
\par
(Department of Physics, Fuzhou University, E-mail: mxc001@163.com)
\end{center}
{\bf Abstracts } It is proved by means of the dynamical effects of special relativity that the velocity caused by accelerating process is not a relative concept. It is an absolute physical quantity that can be determined by experiments. Therefore, the influence of accelerating process must be considered in space-time theory. Besides the Newtonian absolute theory with invariable space-time scales and the Einstein몶s relative theory with variable space-time scales, there exists the third space-time theory, i.e., the absolute space-time theory with variable space-time scales. At present, special relativity is divided into two parts, kinematics and dynamics. In kinematics force and acceleration are not considered so that space and time becomes relative concepts. This division is completely artificial and irrational. The space-time theory should be established on the foundation of dynamics consistently. As the three most fundamental physical quantities, the changing rules of space, time and mass with speed are dependent of accelerating processes and absolute. The real meaning of the invariability principle of high speed should be that we can not make the speed of an object with rest mass reach and exceed the sped of light in vacuum by the method of accelerating it. But if velocity is not caused by accelerating process, it is allowed to exceed light몶s speed in vacuum. For example, apparent tangent velocities of all fixed stars in sky are over light몶s speed. It is proved from the angle of symmetry that the absolutely resting reference frame exists. In order to eliminate multifarious space-time paradoxes in special relativity, and coincide with Big-bang cosmology, the absolutely resting frame is necessary. A method is put forward to look for absolutely resting reference frame. The isotropy reference frame of cosmic microwave background radiation can be regarded as the absolutely resting reference frame actually at present. This absolute space-time theory with variable space-time scales is logically consistent without contradicting any physical experiment and astronomical observation, so it can be considered as a really rational space-time theory to describe real physics world.
\\
{\bf PACS numbers: 03.30.+p, 04.20.Cv, 95.86.Bh, 98.80.Es}\\
{\bf Key Words: Special relativity, Invariability principle of light몶s speed, Lorentz contraction, microwave radiation, Nonequivalent inertial reference frame, Absolutely resting reference frame }\\
\\
\\
{\bf 1. Possibility and necessity of the third space-time theory }\\
\par
모모As we known that according to the Newtonian theory, the scales of space-time are absolute and invariable밃independent of the selection of reference frames. The absolutely resting reference frame was considered to exist, though Newton did not know how to determine it. According the Einstein몶s theory, reversely, the scales of space-time are relative and variable, dependent of the selection of reference frames. The absolutely resting reference frame was considered not to exist. But velocity was always considered as a relative concept whether in Newtonian or in Einstein몶s theories. However, from the angle of logical possibility, the third space-time theory is allowed to exist besides the Newtonian and Einstein몶s theories. That is the absolute space-time theory with variable space-time scales. In this theory, the scales of pace-time are variable but with absolute significance, independent of the selection of reference frames. Besides, velocity caused by acceleration process can be proved to be an absolute concept by means of the dynamic effect of special relativity. This result is completely different from the Newtonian and Einstein몶s theory, and leads to the existence of the absolutely resting reference frame.
\par
모모Lots of experiments show that the scales of space-time are variable, so the Newtonian space-time theory is considered incorrect at present. Though the Einstein몶s theory has obtained great success, multifarious space-time paradoxes can be deduced from it. For example, the Ehrenfest몶s paradox, the Polevaulter몶s paradox, the apparent shape rapidly moving object of paradox, the stress effect paradox due to length contraction, the upper limit paradox on the proper length , the thread paradox, the right-angled lever paradox, the superluminal velocity paradox, the Oppenheimer paradox and other recently raised paradoxes and so on $^{(1)}$. But many persons denied the existence of space-time paradoxes at present. They persists on that those paradoxes were caused only by incorrect understanding of relativity, and put forward many methods to explain the paradoxes, most of them are ambiguous and even completely wrong. For example, in order to explain the paradox of slide falling, the shape of slide was considered different in different inertial reference frames. In the resting frame, the shape of slide is considered to be flat, but when transformed into the moving frame, the shape of slide is thought becoming parabola몶s shape $^{(2)}$. This explanation is completely wrong and unacceptable. An object with flat shape in an inertial frame would still be the same shape in another inertial frame. Otherwise the principle of relativity would be violated. As for the most famous twin paradox with the existence of accelerating processes, the general theory of relativity or equivalent principle has to be used to explain it. This paradox is considered to be solved completely at present $^{(3)}$. However, the problem is not so simple. If we take more symmetrical form, this paradox is impossible to solve. For example, we send two twins into space by two rockets at opposite directions simultaneously, and then let them return to the earth at the same time after arbitrary period of time몶s travel. Because the situations are completely symmetrical for two twins, each twin would think that his brother is younger than himself if time contraction is relative. In this completely symmetrical case, general theory of relativity is also incapable. In fact, as we known, general theory of relativity itself also demands a certain extent of absoluteness. The situation hints us actually that Einstein몶s completely relative space-time theory can not be a final one. It is unsuitable for us to use such a theory with so many paradoxes to describe the real world.
\par
모모It is commonly thought that Einstein몶s space-time theory has nothing to do with acceleration. The theory only considers the measurement relation of space-time between two inertial reference frames. In special relativity, two reference frames are hypothesized to be at rest each other at beginning so that same space-time units (ruler and clock) can be defined. Then by introducing a relative speed between two inertial frames, we can deduce the Lorentz transformation formula based on the principles of relativity and invariability of light몶s speed. From the Lorentz transformation, we can reach a result that space-time몶s contractions are completely relative. However, we meet an inevitable problem here, that is, when a relative speed is introduced into two frames that are at rest at beginning, acceleration process is needed. At rest one of two frames has to be accelerated. Thus, would accelerating process affect space-time characteristics? According to the current understanding at present, it is thought that the effect of accelerating process can be neglected in special relativity. The Lorentz transformation is only relative to speed and speed is considered only a relative concept.
\par
모모However, in order to eliminate time paradox in special relativity, we have to consider the effect of accelerating or non-inertial process on space-time몶s nature. According to equivalent principle, inertial force is equivalent to gravitation, or non-inertial reference frame is locally equivalent to gravitational field. Because space-time scale would change in gravitational field, acceleration process would affect space-time몶s nature if equivalent principle holds. In this way밃we meet a logical absurdity. On the one hand, in order to keep space-time relativity, we have to suppose that accelerating process does not affect space-time몶s nature. On the other hand, in order to reach a proper gravitational theory, we have to suppose that accelerating process is equivalent to gravitation and gravitation field would affect space-time몶s nature. This kind of space-time theory is deviant. It tells us actually that a rational space-time theory can not avoid non-inertial motion몶s effects. Space-time몶s scales would be connected with accelerating process (speaking strictly, both acceleration and accelerating time). 
\par
모모Einstein몶s special relativity is divided into two parts. One is kinematics independent of force and acceleration. Another is dynamics relative to force and acceleration. In kinematics, various paradoxes appear owing to velocity몶s relativity. In dynamics, there is no paradox because force and accelerating process are not relative. Because the effect of accelerating process is neglected, so many paradoxes appear in kinematics. It is factitious and irrational to divided special relativity into two parts. Non-inertial process몶s effect should be considered in rational space-time theory. We should establish space-time theory consolidate on the foundation of dynamics. Only when the dynamic theory to describer space-time몶s nature is established independently, we can consider the relation between non-inertial motion and gravitation. Otherwise the problems would entangle each other before non-inertial motion몶s influence on space-time몶s nature has been known. 
\par
모모Therefore, we have to consider the third space-time theory with absoluteness and variable scales. Velocity caused by accelerating process is proved to be an absolute concept and the absolutely resting frame is demanded to exist. This result coincides with modern cosmology. In fact, we can take the reference frame which was at rest together with the original point of Big-bang universe as the absolutely resting reference frame. All other reference frames in universe are considered to move relative to it. It seems absurd to say that velocity is absolute for it seems to contradict with common knowledge and basic idea of physics. But we can prove it by the effect of special relativity, though it is impossible in Newtonian theory. Only when the effect of special relativity is taken into account, can we say velocity caused by accelerating process to be absolute. We now do it by means of an idea experiment below.
\\
\\
{\bf 2. Absoluteness and anisotropy of velocity caused by accelerating process }\\
\par
모모As shown in Fig.1, suppose there are three resting reference frames represented by $K_0$, $K_1$ and $K_2$ in vacuum in beginning. $K_1$ and $K_2$ are closed chambers. There are two objects $A_1$ and $A_2$ with same masses in both chambers individually.  Each object is connected into the left and right sides of chamber by two elastic ropes. The nature and length of elastic ropes are the same. Then we accelerate $K_2$ in a constant acceleration $a$ toward to right side, while $K_0$ and $K_1$ still keep at rest. After a long enough time, the speed of $K_2$ reaches $V_2=V몶$. According to the dynamical theory of special relativity, inertial mass of object $A_2$ and inertial force acting on elastic rope caused by $A_2$ are individually
\begin{equation}
m_2={{m_0}\over{\sqrt{1-V몶^2/c^2}}}>m_0~~~~~~~~~~~~~~F_2={{m_0{a}}\over{(1-V몶^2/c^2)^{3/2}}}
\end{equation}
Therefore, the elastic rope on the right side of object $A_2$ would be pulled longer and longer in accelerating process. Suppose the biggest force that the elastic rope can bear is $F_2$, if $V_2>V몶$밃the elastic rope would be pulled apart. Suppose the accelerating process is stopped when the speed of $K_2$ just reaches $V_2=V몶$, so the elastic rope has not yet pulled apart in this case. After that, frame $K_2$ moves in a uniform speed $V몶$ towards right side, and elastic rope resumes to original length for it is not acted by the force (Lorentz contraction has not been drawn out for it is unimportant in this discussion.). 
\par
모모Because $K_2$ is in inertial state again, according to space-time relativity, observers in $K_2$ would think that their reference frame is at rest, while $K_0$ and $K_1$reference frames move in uniform speed $V몶$ towards left side. So according to observers in $K_2$, the mass of object $A_2$ is $m_0$ and the moving mass of object $A_1$ is $m_1=m_0/\sqrt{1-V몶^2/c^2}>m_0$. This opinion is just opposite to that of observers in $K_0$ and $K_1$. In the viewpoint of observers in $K_0$ and $K_1$, $K_1$ is at rest in all time and $K_2$ moves towards right side in uniform speed $V몶$. So the mass of object $A_1$ is $m_0$ and the moving mass of object $A_2$ is $m_2>m_0$. Therefore, mass becomes a relative concept for observers in two reference frames with a relative speed between them. 
\par
모모However, it can be proved that this relativity idea is actually impossible. When we want to know an object몶s mass, we need to accelerate it. Then we can decide object몶s mass by the relation of force and acceleration. As shown in Fig.2, we accelerate both $K_1$ and $K_2$ in a constant acceleration $a$ towards right side for a short simultaneously. After $K_1$ reaches a small speed $V_1(0<V_1<<V몶)$, accelerating process is stopped and $K_1$ and $K_2$ resume to the states of uniform motions again. After that, relative to observers in resting frame $K_0$, the speed of $K_2$ becomes $V_1+V_2>V몶$, so that the elastic rope connected to the right side of object $A_2$ would be pulled apart immediately, for the inertial force acted on the rope is bigger than $F_2$. On the other hand, because $K_1$ is only accelerated for a short time and its speed is far small than $V몶$, the elastic rope connected to the right side of object $A_1$ would not be pulled apart. Because whether elastic ropes are pulled apart or not are absolute events, observers in $K_1$ and $K_2$ would observe the same phenomena and admit these facts.
\\
\\
\\
\\
\\
\\
\\
\\
\\
\par                                         
Fig. 1
\\
\\
\\
\\
\\
\\
\\
\\
\par
Fig. 2
\\
\\
\par
모모Thus, we immediately see that the things happen in two inertial reference frames with a relative uniform speed between them are completely different. We can let observers in $K_1$ and $K_2$ are in the state of dormancy from beginning to that $K_2$ reaches uniform speed $V몶$, then awake them. In this case, the observers in two reference frames do not known which one has been accelerated. After that, we accelerate $K_1$ and $K_2$ for a while again. By means of the facts that the rope connected to object $A_2$ is pulled apart, but the rope connected to object $A_1$ is not pulled apart, observers in both reference frames can obtain the same viewpoint that $K_2$ is accelerated even and obtains a speed $V몶$. Meanwhile, $K_1$ is at rest before the second accelerating process is carried out. There is no relativity in this case. Velocity caused by accelerating process becomes an absolute effect. It is obvious that this result violates the principle of space-time relativity in special relativity. 
\par
모모As we know that in Newtonian theory, object몶s mass does not change with speed, so rope would not be pulled apart in the process above. It means that we can not decide which reference frame moves and which is at rest by the same experiment according to the Newtonian theory. So in the Newtonian theory, velocity is still a relative concept. It is just by means of the dynamical effect of special relativity, the velocity produced by accelerating process becomes an absolute physical quantity. 
\par
모모Therefore밃moving mass of a object should be also an absolute physical quantity. When there exists a relative speed between $K_1$ and $K_2$, observers in both reference frames would have the same viewpoint about the masses of objects $A_1$ and $A_2$. Because $K_2$ is accelerated, the mass of object $A_2$ increase. Because $K_1$ is at rest, the mass of object $A_1$ is unchanged. There is no relativity here. The accelerating process has an effect on an object몶s mass. This is easy to understanding from the angle of energy conservation. Energy is needed when a force is acted on an object and work is done. The result is that object몶s energy or motion mass increases. Therefore, it is different from traditional idea that the motion of an object has reason. The reason comes from force and acceleration. Some observable and measurable results would be caused when the object obtain a velocity through accelerating process. Unfortunately in the Newtonian and Einstein몶s space-time theories, this point is neglected so that velocity becomes something that can be designated arbitrary without origin. 
\par
모모Now let us to show that velocity caused by accelerating process is anisotropic in general. For this purpose, let몶s discuss the process of decelerating $K_2$. As shown in Fig.3, when $K_2$ moves in a uniform speed $V몶$ without rope being pulled apart, we accelerate $K_2$ towards left side in same acceleration $a$ (actually decrease $K_2$). At beginning, the velocity of $K_2$ is still towards to right side. After a period of time, the speed of $K_2$ is decreased to zero, so that three reference frames become at rest each other again. Because in whole process the speed of $K_2$ is always with $V_2<V몶$, the rope connected to the left side of object $A_2$ would not be pulled apart. Because whether the rope is pulled apart or not is an absolute event, observers in $K_2$ would also observer the same result and hold the same viewpoint.
\\
\\
\\
\\
\\
\\
\\
\\
\par
Fig.3
\\
\\
\par
모모Thus we can see immediately that the results are completely different in both cases that $K_2$ is accelerated towards right side and towards left side. When it is accelerated towards right side, the rope would be pulled apart. But when it is accelerated towards left side, the rope would not be pulled apart. All of observers in three frames agree with this result. So observers in reference frame $K_2$ would think that space is with anisotropy for him, because the results are different when his reference frame is accelerated towards different directions. This result also violates the principle of space-time relativity in special relativity.
\\
\\
{\bf 3. Nonequivalent inertial frames and absoluteness of Lorentz contractions }\\
\par
모모So we have to consider the effect of accelerating process on space-time몶s characteristics. Based on the consideration of accelerating process, we can establish a more rational space-time theory. At first, let몶s introduce the concept of nonequivalent inertial reference frames. We call such two reference frames as nonequivalent inertial reference frames. At beginning, these two reference frames are at rest each other and define the same unit scales of space and time (ruler and clock). Then by accelerating one of them at least, a motion velocity is introduced between them. After accelerating process is stopped, they become inertial reference frames again with a relative velocity between them. That is to say, the relative velocity between two nonequivalent inertial frames is produced by accelerating process.  
\par
모모In fact, all of human몶s common activities and scientific experiments are carried out on nonequivalent reference frames. Most of measurements about the relation of space and time are done in nonequivalent reference frames. For example, Michelson interferometer몶s two arms are just nonequivalent reference frames when we use the interferometer to verify the invariability principle of light몶s speed. When one of arms moves along with the earth몶s moving direction, another arm can be considered at rest. When the interferometer rotates, the motion states of two arms exchange each other throuth non-inertial rotating. In fact, the earth is just a big non-inertial system, all scientific experiments done on it involves the effects of non-internal motion. Only based on nonequivalent reference frames, the theory is meaningful. Once theory is established on nonequivalent reference frames, absoluteness and asymmetry would be introduced. 
\par
모모Therefore, two foundational hypothesis of Einstein몶s special relativity should be revised as follows 
\par
1밅 Non-inertial motions would affect the space-time몶s natures of non-inertial reference frames. But the description forms of physics laws are the same in non-inertial reference frames.\\
\par
2밅 The speed of light in vacuum is invariable among arbitrary non-inertial reference frames. \\
\par
모모By means of these two hypotheses, the same Lorentz transformation can also be deduced, but the transformation is with absolute significance now. In order to obtain the transformations, we suppose that there are two reference frames $K_1$ and $K_2$ in vacuum. In the beginning they are at rest each other and the same rulers and clocks are defined. Then let $K_1$ is still at rest and $K_2$ is accelerated towards right side. When the speed of $K_2$ reaches $V$, accelerating process is stopped. After that, $K_2$ moves in a uniform speed relative to $K_1$. Suppose at a certain moment, point $x몶_1$ at $K_1$ meets point $x몶_2$ at $K_2$. At this time, the clock at point $x몶_1$ indicates $t몶_1$, and the clock at point $x몶_2$ indicates $t몶_2$. Meanwhile, a bind of light is send out along the $X$ axes of two frames. This light reaches the point $x_1$ at time $t_1$ relative to $K_1$, reaches the point $x_2$ at time $t_2$ relative to $K_2$. In light of the invariability principle of light몶s speed between non-inertial reference frames, we have
\begin{equation}
(x_1-x몶_1)^2-c^2(t_1-t몶_1)^2=0~~~~~~~~~~~~~~ (x_2-x몶_2)^2-c^2(t_2-t몶_2)^2=0
\end{equation}
From these two equations, we can deduced the Lorentz formulas 
\begin{equation}
x_2-x몶_2={{x_1-x몶_1-V(t_1-t몶_1)}\over{\sqrt{1-V^2/c^2}}}~~~~~~~~~~~~~~t_2-t몶_2={{t_1-t몶_1-V(x_1-x몶_1)/c^2}\over{\sqrt{1-V^2/c^2}}}
\end{equation}

Suppose space and time intervals for two reference frames are $\triangle{l}_1=x_1-x몶_1$, $\triangle{l}_2=x_2-x몶_2$, $\triangle{t}_1=t_1-t몶_1$ and $\triangle{t}_2=t_2-t몶_2$ individually밃we can obtain the same Lorentz contraction formulas
\begin{equation}
\triangle{l}_1=\triangle{l}_2\sqrt{1-{{V^2}\over{c^2}}}~~~~~~~~~~~~~\triangle{t}_1={{\triangle{t}_2}\over{\sqrt{1-V^2/c^2}}}
\end{equation}

It shows that length becomes short and time becomes slow for reference frame $K_2$ moving in uniform speed $V$. Because the motion speed of $K_2$ is caused by accelerating process, we should think that the real change of space-time몶s scales of $K_2$ reference frame takes place owing to the effect of non-inertial process. Though the formulas of space-time contractions are relative to speed, it is only a superficies. The effects of accelerating process are hided. When accelerating process stops, space-time몶s scales are fixed. 
\par
모모On the other hand, we let $K_2$ at rest and accelerate $K_1$ towards left side in the beginning. When speed of $K_2$ reaches $-V$, stop accelerating process, we obtain the conversed Lorentz transformation
\begin{equation}
x_1-x몶_1={{x_2-x몶_2+V(t_2-t몶_2)}\over{\sqrt{1-V^2/c^2}}}~~~~~~~~~t_1-t몶_1={{t_2-t몶_2+V(x_2-x몶_2)/c^2}\over{\sqrt{1-V^2/c^2}}}
\end{equation}

As well as the conversed space-time contraction formulas
\begin{equation}
\triangle{l}_2=\triangle{l}_1\sqrt{1-{{V^2}\over{c^2}}}~~~~~~~~~~~~~~~~\triangle{t}_2={{\triangle{t}_1}\over{\sqrt{1-V^2/c^2}}}
\end{equation}

\par
모모It means that the ruler become short and clock becomes slow for frame $K_1$ moving in velocity $-V$. Similarly, the changes of space-time몶s scales caused by accelerating process are also real and absolute. Which reference frame is accelerated, of which space-time scales are changed. So it can be seen that the premises of Lorentz transformation formulas (3) and (5) are different. In fact, before of Einstein몶s relativity explanation, Lorentz thought that space-time몶s contraction was absolute effect owing to the motions of objects relative to resting aether. We are used to regard reference frames as some things which are abstract and arbitrary velocity can be endowed to. However, any practical reference frame is always composed of material with mass. Force or interaction is always needed to accelerate a practical reference frame. But this fact is always omitted in space-time theory. 
\par
모모In order to explain the relativity of length contraction, Einstein put forwarded the concept of relativity of simultaneity. The concept claims that if two events at two different sites of an inertial reference frame take place simultaneously, they would not take place simultaneously for another inertial reference frame with a relative speed. In this way, the relativity of length contraction is turned over to the relativity of simultaneity. However, for nonequivalent inertial reference frames, there is no the relativity of simultaneity. Though the readings of clocks located at the different places of an accelerated reference frames are different, the concept of simultaneity is still absolute.  For nonequivalent inertial frames, if two events take place simultaneously for a reference frame, they also take place simultaneously for another reference frame, though the time몶s readings on the different clocks of two reference frames may be different.
\par
모모On the other hand, according to this absolute and variable space-time theory, the real significance of the invariability principle of light몶s speed is that we can not let the speed of an object with rest mass reach and exceed light몶s speed in vacuum by the method of accelerating it. Only in this meaning, we can consider light몶s speed as a limit speed.
\\
\\
{\bf 4. Necessity and possibility of absolute resting reference frame}\\
\par
     As shown in Fig.4, let $K_0$ represent absolute resting reference frame. $K_1$, $K_2$ and $K_0$ are at rest in beginning and same rulers and clocks are defined for them. Then we accelerate $K_1$ and $K_2$ towards right sides in accelerations $a_1$ and $a_2$ with $a_2>a_1$ individually and simultaneously. When the speed of $K_1$ reaches $V_1$ and the speed of $K_2$ reaches $V_2$, stopping accelerating processes. After that, $K_1$ and $K_2$ move relative to $K_0$ in uniform in uniform speeds $V_1$ and $V_2$ with $V_2>V_1$ individually. Suppose at a certain moment, the original points of three reference frame몶s coordinates just meet together. The readings of clocks at three frame몶s original points are adjusted to $t_0=t_1=t_2=0$ at this time. Meanwhile, a bind of light is send out along the $X$ axes. In light of the invariability principle of light몶 speed of non-equivalent inertial frames, the Lorentz transformations can be written as
\begin{equation}
x_1={{x_0-V_1{t}_0}\over{\sqrt{1-V^2_1/c^2}}}~~~~~~~~~~~~~~~t_1={{t_0-V_1{x}_0/c^2}\over{\sqrt{1-V^2_1/c^2}}}
\end{equation}

\begin{equation}
x_2={{x_0-V_2{t}_0}\over{\sqrt{1-V^2_2/c^2}}}~~~~~~~~~~~~~~t_2={{t_0-V_2{x}_0/c^2}\over{\sqrt{1-V^2_2/c^2}}}
\end{equation}

The relative speed between $K_1$ and $K_2$ is
\begin{equation}
V={{V_2-V_1}\over{1-V_1{V}_2/c^2}}
\end{equation}

The space-time transformation relations between $K_1$ and $K_2$ are
\begin{equation}
x_2={{x_1-Vt_1}\over{\sqrt{1-V^2/c^2}}}~~~~~~~~~~~~~~t_2={{t_1-Vx_1/c^2}\over{\sqrt{1-V^2/c^2}}}
\end{equation}

The invariability principle of light몶s speed between $K_1$ and $K_2$ still holds, but their relative speed should be defined by Eq.(9). Because of $a_2>a_1$, $V_2>V_1$ in this case, we should think that $K_2$ is accelerated relative to $K_1$, so that the ruler and clock of $K_2$ contract absolutely relative to $K_1$. 
\\
\\
\\
\\
\\
\\
\\
\\
\par
Fig. 4
\\
\\
\par
모모On the other hand, as shown in Eq.(7), when accelerated $K_1$ moves in uniform speed $V_1$, the clocks located at different places dedicate different times. The observers in $K_1$ can regulate these clocks so that their time becomes the same. After that, the observers in $K_1$ would find that light몶s speed is different from $K_0$. There is no light몶s speed invariability again between $K_0$ and $K_1$ in this case. If the observers in $K_1$ accelerate another reference frame $K몶_1$ with same ruler and clock. Then stop accelerating when $K몶_1$ reaches a certain speed. After that, there exists the invariability of light몶s speed and the Lorentz transformation between $K_1$ and $K몶_1$, but there exists no the invariability of light몶s speed and the Lorentz transformation between $K몶_1$ and $K_0$ (as well as $K몶_1$ and $K_2$). In fact, all nonequivalent inertial reference frames on the earth are established in this way. 
\par
모모The discussion above would lead to a result, that is, the absolutely resting reference frame exists. In fact as shown in Fig.4, if we accelerate $K_1$ towards right side so that its speed becomes faster after it obtain a uniform speed $V_1$, the rulers in $K_1$ would become shorter further, the clocks would become slower further and the mass of object would becomes greater further. Oppositely, if we accelerate $K_1$ towards left side so that its speed becomes slower after it obtains a uniform speed $V_1$, the rulers in $K_1$ would become longer and the clocks would become quicker and the masses of objects would become smaller. When the speed of $K_1$ becomes zero relative to the absolutely resting reference frame $K_0$, the rulers are longest, the clocks are quickest and the masses of objects are smallest. If we continue to accelerate $K_1$ towards left side, the rulers in $K_1$ would become short and the clocks would become slow and the masses of objects would become great again. 
\par
모모It is obvious that only the space-time of absolutely resting reference frame is isotropy. The rulers are longest, the clocks are quickest and masses are smallest in the absolutely resting reference frame. No other frames have such characteristics. So from the angle of symmetry, we can always find out such a reference frame, in which space-time is completely isotropy. We can do it by accelerating a reference frame in different directions in space, and observer the changes of ruler, clock and object몶s mass in the frame. If experimental results are isotropy at a certain state, the reference frame in this case can be regarded as the absolutely resting reference frame. 
\par
모모A simple method is provided here to determine the absolutely resting reference frame. Because object몶s moving mass is the smallest in the absolutely resting reference frame (equal to its rest mass), as long as we find out a reference frame in which a object몶s mass is smallest, we can think that this reference frame is just the absolutely resting one. The concrete method is below. A chamber such as shown in Fig. 1.1 is fixed in a rocket. An object is connected to a sensitive spring. Then we accelerate the rocket in a constant acceleration in different directions in space. The spring would be pulled long by the inertial force caused by the object. Measure the length몶s change of spring accurately by optical method. If we find that the length of spring is the shortest at a certain moment, stopping accelerating rocket. Then the rocket would move in a uniform speed in space. In this case, all inertial reference frames which are at rest with the rockets can be regarded as the absolutely resting reference frames. 
\par
모모The existence of absolutely resting reference frame coincides with the demand of modern cosmology and practical observations of astronomy. In the 1960몶s, astronomers found the special anisotropy of microwave background radiation $^{(4)}$. If the isotropic reference frame of microwave background radiation is taken as the absolutely resting one, astronomic observations shown that earth frame was moving in a speed of $V\leq{300}$ Meters/s towards to the direction of right ascension $0\sim{13}^{h}$. This velocity could be regarded as the velocity that the earth moves relative to the absolutely resting reference frame. In 1999, the anisotropy detector of microwave background radiation (WMAP) found anisotropy at higher precision. In 2002, two physicists in Oxford University found the anisotropy of radio waves eradiated by radio galaxy in the earth몶s motion direction by using array radio telescopes (VLA) $^{(5)}$. This kind of anisotropy is considered to be caused by the Doppler effect of the earth몶s motion. Any more, the Big-bang cosmology demands the existence of the absolutely resting reference frame. We can take the reference frame which is at rest with the original point of Big-bang universe as the absolutely resting reference frame. The other reference frames produced by accelerating in the Big-bang processes would be considered to move relative it. It seems at present that the isotropic reference frame of microwave background radiation may be taken as the absolute resting one. 
\\
\\
{\bf 5. Equivalent inertial reference frames and possibility of over light speed몶s motion }\\
\par
모모We define reference frames as equivalent inertial reference frames that relative velocities between them exist primordially, not be caused by accelerating processes. That is to say, there exist no relative resting states between equivalent inertial frames from beginning to end, so that they can not define same rulers and clocks at resting states. The definitions of rulers and clocks in equivalent inertial frames can be independent. These kinds of inertial reference frames, of cause, are completely equivalent. Because such equivalent inertial frames may have no any causality, they are not restricted by the invariability principle of light몶s speed in vacuum. In fact, all moving reference frames produced by the Big-bang processes can not be equivalent inertial frames, and relative moving velocities between them can not pass over light몶s speed in vacuum. But if there are some celestial bodies which were not produced through the Big Bang processes of our universe, relative velocities between our universal celestial bodies and these celestial bodies can be allowed to pass over light몶s velocity in vacuum. 
\par
모모On the other hand, for non-inertial reference frames, some apparent relative velocities may be allowed to pass over light몶s velocity when these velocities are not caused by accelerating processes. For example, the tangent velocities of all fixed stars in space exceed light몶s speed observed by observers on the rotating earth. But no observers on the earth find that the fixed star몶s diameters become zero even imaginary number. If we build a uniformly rotating disc on the earth, as long as the angle speed of disc reaches $\omega\geq{1/480}$, the tangent speed of the sum would reach and pass over light몶s speed. But the diameter of the sum is almost unchanged for the observers on the disc. All these facts can not be explained by Einstein몶s special relativity. But it can be explained well by this paper몶s theory. This is because that the sum and fixed stars have not ever been accelerated, so their diameters do not contract. It is that the rotating disc is accelerated ever, so it is that tangent length of disc contracts. So we only can say that the rotating earth and disc are moving, instead of the sun and fixed stars. By using the ruler in the rotating disc to measure, the diameters of the sum and fixed stars would become longer, instead of shorter, for the ruler becomes shorter. Though the change is very small when the tangent speed of disc is very small with $V=r\omega<<c$. Meanwhile, all observers on the earth, sun or fixed stars have the same viewpoint on motion and velocity, there is no relativity. In fact, energy needed to accelerate a fixed star is completely different from that to accelerate a disc. This kind of asymmetry would cause some measurable effects. 
\par
모모It should be emphasized again that the invariability principle of light몶s speed only means that the speed of an object with rest mass can not reach and exceed light몶s speed in vacuum by means of the method of accelerating it. The reason to cause light몶s speed invariable is that the space-time scales of non-inertial reference frame change really in accelerating process, so that the observers in the non-inertial frame find that light몶s speed is still unchanged. In fact, so-called light몶s speed invariability is conditional. For example, in medium, light몶s speed is variable. It depends on medium몶s refractive index. When refractive index is big than 1, the speed of light in medium is smaller than its speed in vacuum. This is a well-known fact in physics. If some day we find that a certain medium몶s refractive index is small than 1, it needs not to surprise that light몶s speed in this medium would pass over its speed in vacuum. In fact, some authors declaimed in 2000 that they had found the phenomena of over-light몶s speed in atom $C_s$ gas $^{(6)}$.
\par
모모In sum, because velocity caused by accelerating process can not be relative, we have to abandon the principle of relativity, while the invariability principle of light몶s speeds between non-inertial reference frames are hold. We should establish space-time transformation theory based on the non-inertial reference frames. In this way, we can reach an absolute space-time theory with variable space-time scales. This kind of theory has no any logical contradiction and confusion. It does not contradict with any present physical experiments and astronomic observations, and also coincides with modern cosmology. So it can be considered as a really rational space-time theory to describe this real physics world.
\\
\\
\\
\\
References
\\
1. Sastry G., Am. J. phys., 55, 943, (1987). Varica V. Zum, Z. Phys, 12, 169 (1911). 
모Ai H B., Phys Essays, 11, 1 (1998). \\
2. W. Rindler, Essential Relativity, Second Edition, (1977).\\
3. Zhang Yongli, Introduce to Relativity, 286밃(1980).\\
4. S. Weiberge, Gravitation and Cosmology, 608밃(1984).\\
5. Nature, 3, (2002).\\
6. Wang, Lijun, et al, Nature, 406, (2000).\\
\end{document}